\documentclass[preprints,article,accept,moreauthors,pdftex]{mdpi} %
\usepackage{color}
\usepackage{subcaption}
\usepackage{graphicx}
\firstpage{1} 
\articlenumber{x}
\doinum{10.3390/------}
\pubvolume{xx}
\pubyear{2017}
\copyrightyear{2017}
\externaleditor{Academic Editor: name}
\history{Received: date; Accepted: date; Published: date}

\Title{Enhanced Efficiency at Maximum Power in a Fock-Darwin Model Quantum Dot Engine}

\Author{Francisco J. Pe\~na  $^{1,2}$*, Nathan M Myers$^{3}$*, Daniel \'Ordenes$^{1,2}$, Francisco Albarr\'an-Arriagada$^{4}$ and Patricio Vargas $^{5}$}

\AuthorNames{N. M. Myers, D. \'Ordenes, F. J. Pe\~na,  F. Albarr\'an-Arriagada and P. Vargas}

\address{%
	$^{1}$ \quad Departamento de F\'isica, Universidad T\'ecnica Federico Santa Mar\'ia, Av. Espa\~na 1680, Valpara\'iso 11520, Chile;  francisco.penar@usm.cl (F.J.P.), daniel.ordenes@usm.cl (D.O.); \\ 
    $^{2}$ \quad Millennium Nucleus in NanoBioPhysics (NNBP), Av. Espa\~na 1680, Valpara\'iso 11520, Chile;  \\ 
    $^{3}$ \quad Department of Physics, Virginia Tech, Blacksburg, Virginia 24061, USA; myersn1@vt.edu (N.M.M.) \\
	$^{4}$  \quad Departamento de F\'isica, CEDENNA, Universidad de Santiago de Chile (USACH), Avenida V\'ictor Jara 3493, 9170124, Estaci\'on Central, Chile. francisco.albarran@usach.cl (F.A.A.)\\
	$^{5}$ \quad Departamento de F\'isica, CEDENNA, Universidad T\'ecnica Federico Santa Mar\'ia,  Av. Espa\~na 1680, Valpara\'iso 11520, Chile;  patricio.vargas@usm.cl (P.V.)}

\corres{Correspondence:  francisco.penar@usm.cl; myersn1@vt.edu}
 
\abstract{We study the performance of an endoreversible magnetic Otto cycle with a working substance composed of a single quantum dot described using the well-known Fock-Darwin model. We find that tuning the intensity of the parabolic trap (geometrical confinement) impacts the proposed cycle's performance, quantified by the power, work, efficiency, and parameter region where the cycle operates as an engine. We demonstrate that a parameter region exists where the efficiency at maximum output power exceeds the Curzon-Ahlborn efficiency, the efficiency at maximum power achieved by a classical working substance.}


\begin{document}
\section{Introduction}

The study of quantum thermal machines, devices such as quantum engines and  refrigerators that operate on working mediums composed of quantum systems, has proved to be a fruitful research area in the last decade~\cite{Kosloff2014AnnuRevPhysChem, Quan2007PhysRevE, Quan2009PhysRevE, Bhattacharjee2021EPJB,Myers2022}. These efforts have been focused principally on the study of the different thermodynamics cycles, such as the Otto and the Stirling cycles, in different regimes~\cite{Chen2019PhysRevE,Myers2021NewJPhys,deAssis2019PhysRevLett,Yin2017EPJPlus,Raja2021NewJPhys} and on the use of different quantum working substances such as spins~\cite{Peterson2019PhysRevLett,Cakmak2019PhysRevE,Ji2022PhysRevLett,Geva1992,Henrich2007}, quantum dots~\cite{Josefsson2018NatNanotechnol,Erdman2017PhysRevB}, and quantum harmonic oscillators~\cite{Kosloff2017Entropy,Deffner2018Entropy,Ding2018,Abah2012,MyersN2019}.  Recent work has also examined the role that quantum properties, such as quantum coherence and quantum correlations, play in the performance of quantum thermal machines ~\cite{Shi2020JPhysAMathTheor,Altintas2015PhysRevA,Rahav2012PhysRevA,Barrios2017PhysRevA,Altintas2014PhysRevE,Park2013PhysRevLett}. A primary feature of interest of quantum thermal machines is their potential to surpass the  performance of their classical counterparts~\cite{Hamman2022arXiv,Jaramillo2016NewJPhys}, which open the door to a new generation of highly efficient quantum engines and refrigerators for application in emerging quantum technologies.

While recent years have seen the experimental implementation of some  quantum thermodynamic cycles~\cite{Peterson2019PhysRevLett, Ji2022PhysRevLett, Rossnagel2016Science,Koch2022}, these devices serve primarily as proof-of-concept prototypes rather than a practical means of extracting useful energy in the form of work. This is since coupling a quantum thermal machine to another device to extract and use the work is a complex and challenging task within the frameworks in which these quantum engines have been implemented. A promising platform for surpassing these difficulties is quantum dots, whose easy fabrication and controllability have led to significant progress in studying their thermoelectric properties \cite{Josefsson2018NatNanotechnol, StaringEPL1993, SvilansCRP2016}. A thermodynamic cycle can be implemented with a quantum dot working medium by manipulating external parameters such as magnetic fields, resulting in an electrical current in the quantum dot during the different steps of the  cycle~\cite{Esposito2010PhysRevE, Sothmann2012EPL, Josefsson2018NatNanotechnol, Pena2019Entropy, Munoz2014PhysRevE}. In principle, this current can be extracted for use in another device with currently available technology. 

In the context of one-electron systems and quantum dots, the famous Fock-Darwin model is very well studied. This model describes an electron in a circular semiconductor quantum dot confined by a parabolic potential under a perpendicular external magnetic field \cite{Jacak1988}. The Fock-Darwin model is fundamental to understanding the behavior of quantum dots under an external magnetic field and demonstrates the competition between geometrical confinement and the confinement induced by the external field \cite{Chen2007}. The Fock-Darwin energy spectra for conventional quantum dots have also been verified experimentally using transport spectroscopy  \cite{Chen2007,McEuen1991,Weis1992,Tarucha1996}.


Thermodynamic endoreversibility corresponds to a non-equilibrium approximation that assumes internally reversible subsystems that then exchange energy irreversibly among themselves \cite{Hoffmann2005}. Thus any dissipation arises purely from the interactions of these subsystems \cite{Andresen1977, Rubin1979, Rubin1979(2), Hoffmann1997, Brown1986}. The common feature of endoreversible engine analysis is an additional limitation on the cycle efficiency caused by the finite rate at which heat can be exchanged between the working substance and the thermal reservoirs. By applying the endoreversible model to the Carnot cycle, Curzon and Ahlborn (CA) demonstrated that the efficiency at maximum power of the classical Carnot cycle is,
\begin{equation}
\label{CAeff}
    \eta_{CA}=1-\sqrt{\frac{T_{c}}{T_{h}}}, 
\end{equation}
where $T_{c}$ and $T_{h}$ are the cold and hot reservoirs temperatures, respectively. Notably, like the Carnot efficiency, the CA efficiency is independent of the nature of the working substance.

Recently, in Ref. \cite{Deffner2018Entropy} Deffner showed that applying the endoreversible approach to the case of an Otto cycle with a single classical or quantum harmonic oscillator as the working medium yields an efficiency at maximum power equal to CA for the classical working medium, and an efficiency at maximum power greater than CA for the quantum working medium. One of the main conclusions of this work is that, unlike the Carnot efficiency, the Curzon-Ahlborn efficiency is not a general upper bound on efficiency at maximum power. That is, it is entirely possible to find  working substances for which the efficiency at maximum power may exceed the CA efficiency, as has been shown to be true in the case of continuous engines operating outside the linear regime \cite{Jordan2013, Benenti2013}.  With this in mind it is of interest to explore whether other quantum working mediums, such as quantum dots, may allow for similar boosts in engine performance.

In this work, we will study a finite-time endoreversible magnetic Otto cycle, using a Fock-Darwin model quantum dot as a working substance. We find that modifying the dot geometry strongly impacts all examined cycle characteristics, including increasing the efficiency at maximum power beyond the CA efficiency. These results indicate that quantum dots can serve as a viable platform of technological interest for implementing endoreversible quantum thermodynamic cycles.

This manuscript is organized as follows. In Sec.~\ref{Sec2}, we introduce the model that describes our working substance, including the energy spectrum, the partition function, entropy, and internal energy. In Sec.~\ref{Sec3}, we describe the thermodynamic cycle under consideration, obtaining the expressions for work, heat, efficiency, and power. In Sec.~\ref{Sec4}, we determine the efficiency at maximum power and identify the conditions under which our system can surpass the CA efficiency. Finally, in Sec.~\ref{Sec5}, we provide a summary of the main conclusions of our work.

\section{Model}
\label{Sec2}

Our working substance is described by the Fock-Darwin (FD) model. The FD Hamiltonian corresponds to a cylindrical quantum dot subject to an external magnetic field perpendicular to the plane in which the dot is present, 
\begin{equation}
    \mathcal{H}= \frac{1}{2m^{*}}\left[\left(p_{x}-\frac{eyB}{2}\right) + \left(p_{y} + \frac{exB}{2}\right)\right] + \frac{1}{2}m^{*}\omega_{0}^{2}\left(x^{2} + y^{2}\right),
\end{equation}
where $e$ is the electron charge, $m^{*}$ is the effective mass of the electron (characteristic of each material), $\omega_0$ is an effective parameter for the strength of the harmonic trap, and $B$ is the intensity of the magnetic field in the $z$ direction. The problem has cylindrical symmetry and eigenenergies,
\begin{eqnarray}
E_{nm}=\hbar\Omega\left(2n+\mid m \mid+1\right)+\frac{1}{2}\hbar\omega_{c}m, 
\label{fockdarwin}
\end{eqnarray}
where $\omega_{c}=eB/m^{*}$ is the cyclotron frequency, $n$ and $m$ are the radial and magnetic quantum numbers ($n$=0,1,2,... and $m$=$-\infty$,...,+$\infty$), respectively, and $\Omega$ is the effective frequency of the system,
\begin{eqnarray}
\Omega=\omega_{0}\left(1+\left(\frac{\omega_{c}}{2\omega_{0}}\right)^{2}\right)^{\frac{1}{2}}.
\end{eqnarray}
Notice that  when the parameter $\omega_{0} \rightarrow 0$, the energy levels of Eq.~(\ref{fockdarwin}) take the usual form of the Landau energy levels in cylindrical coordinates. For high magnetic fields, $(\omega_{c}/2\omega_{0}>> 1)$, Eq.~(\ref{fockdarwin}) simplifies to,
\begin{equation}
E_{n,m}= \frac{\hbar \omega_{c}}{2} (n + 1/2 +| m | + m ) \;,
\label{highfield}
\end{equation}
as $|m|+m  = 0$ for $m<0$. We note that each Landau level, labeled by $n$, is infinitely degenerate in this limit.



In this paper, we consider a low-frequency coupling for the parabolic trap given by $\omega_{0}\sim 2.637$ THz, which in terms of energy units corresponds to a coupling of approximately 1.7 meV. This value is comparable to the typical energy of intra-band optical transitions of quantum dots \cite{Jacak1988}. The order of this transition is approximately $\sim 1$ meV  for cylindrical GaAs quantum dots with effective mass $m^{*}\sim 0.067 m_{e}$ \cite{Jacak1988,Munoz2005,Mani2002}. The quantum dot length can be calculated using $l_{dot}=\sqrt{\hbar/m^{*}\omega_{0}}\sim 25$ nm for the above values. The effective mass gives a cyclotron frequency of $\omega_{c}(B=1)\sim 2.62$ THz for an intensity of $B=1$ T. It is important to mention that we will neglect Zeeman splitting and spin-orbit interaction, which is very small for the GaAs systems \cite{Jacak1988,Munoz2005}.
 
The partition function for our working substance can be written as ~\cite{KumarPhysRevE2009},
\begin{equation}
\mathcal{Z}_{d}=\frac{\omega_{+}\omega_{-}}{4\omega_{0}^{2}}\text{csch}\left(\frac{\hbar\beta\omega_{+}}{2}\right)\text{csch}\left(\frac{\hbar\beta\omega_{-}}{2}\right),
\label{partitionfunctionspin}
\end{equation}
where $\beta = 1/k_{B} T$ is the inverse temperature and the frequencies $\omega_{\pm}$  are given by,
\begin{equation}
\label{frequencies}
\omega_{\pm}^{2}=\frac{1}{2} \left(\omega_{c}^{2} + 2\omega_{0}^{2}\pm \omega_{c}\sqrt{\omega_{c}^{2}+\omega_{0}^{2}}\right).
\end{equation}
Note that if $\omega_{0}\rightarrow 0 $, then $\omega_{-}\rightarrow 0$ and $\omega_{+}\rightarrow \omega_{c}$, recovering the typical partition function for the Landau problem. 

The entropy, $S(T, B)$, and internal energy, $U(T, B)$, are derived from the partition function using, 
\begin{equation}
S(T,B)=k_{B}\ln\mathcal{Z}_{d} + k_{B}T\left(\frac{\partial \ln \mathcal{Z}_{d}}{\partial T}\right)_{B},
\label{eq_s}
\end{equation}
and,
\begin{equation}
U(T,B)=k_{B}T^{2}\left(\frac{\partial \ln\mathcal{Z}_{d}}{\partial T}\right)_{B}.
\label{eq_u}
\end{equation}
In particular, the internal energy is expressed entirely in terms of $\beta$, $\omega_{+}$, and $\omega_{-}$,
\begin{equation}
    U(T,B)= \frac{1}{2}\left[\hbar \omega_{+} \coth\left(\frac{\hbar\beta\omega_{+}}{2}\right) + \hbar \omega_{-} \coth\left(\frac{\hbar\beta\omega_{-}}{2}\right)\right].
\end{equation}
Note that this expression is simply the sum of the internal energies of two oscillators, one with frequency $\omega_+$ and  the other with frequency $\omega_-$. Conversely, the entropy cannot be decomposed in this manner due to the prefactor present in the partition function in Eq.~(\ref{partitionfunctionspin}). This prefactor accounts for the degeneracy of the energy levels and depends on the external magnetic field. Only in the absence of this term can the entropy be expressed as the sum of the entropies of two independent oscillators with frequencies $\omega_{+}$ and $\omega_{-}$.

\begin{figure}
\begin{subfigure}{.5\textwidth}
  \centering
  \includegraphics[width=.8\linewidth]{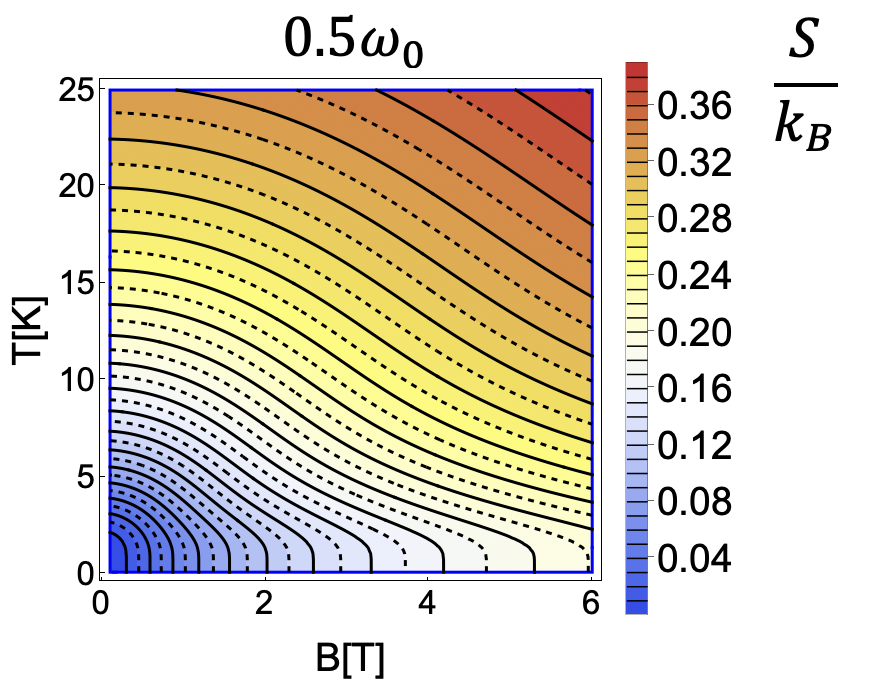}
  \caption{}
  \label{fig:sfig1}
\end{subfigure}
\begin{subfigure}{.5\textwidth}
  \centering
  \includegraphics[width=0.8\linewidth]{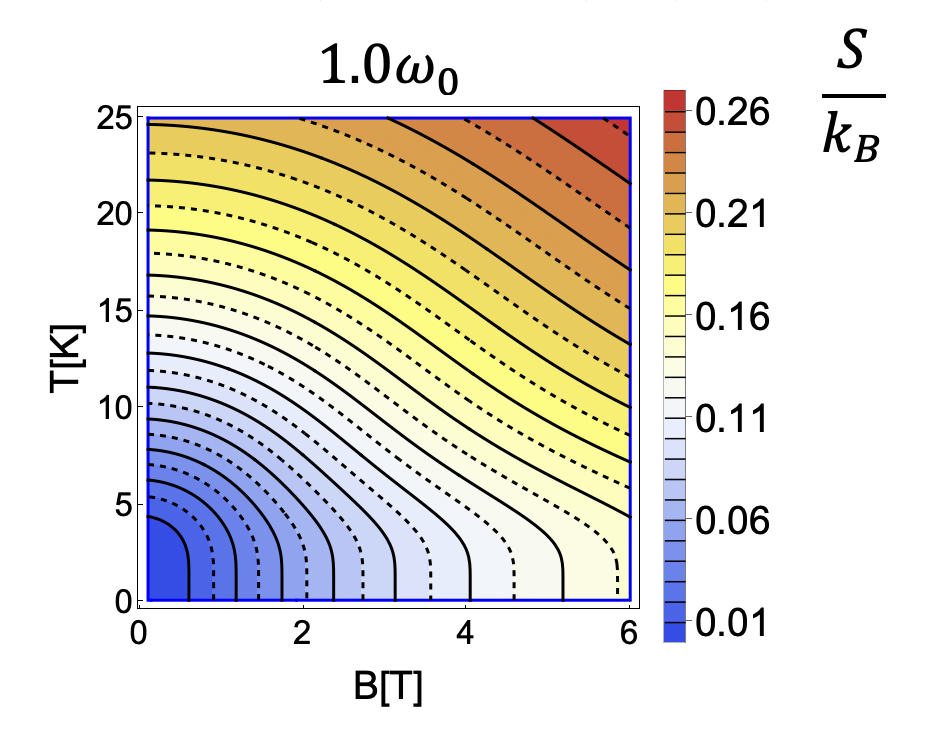}
  \caption{}
  \label{fig:sfig2}
\end{subfigure}
\begin{subfigure}{.5\textwidth}
  \centering
  \includegraphics[width=0.8\linewidth]{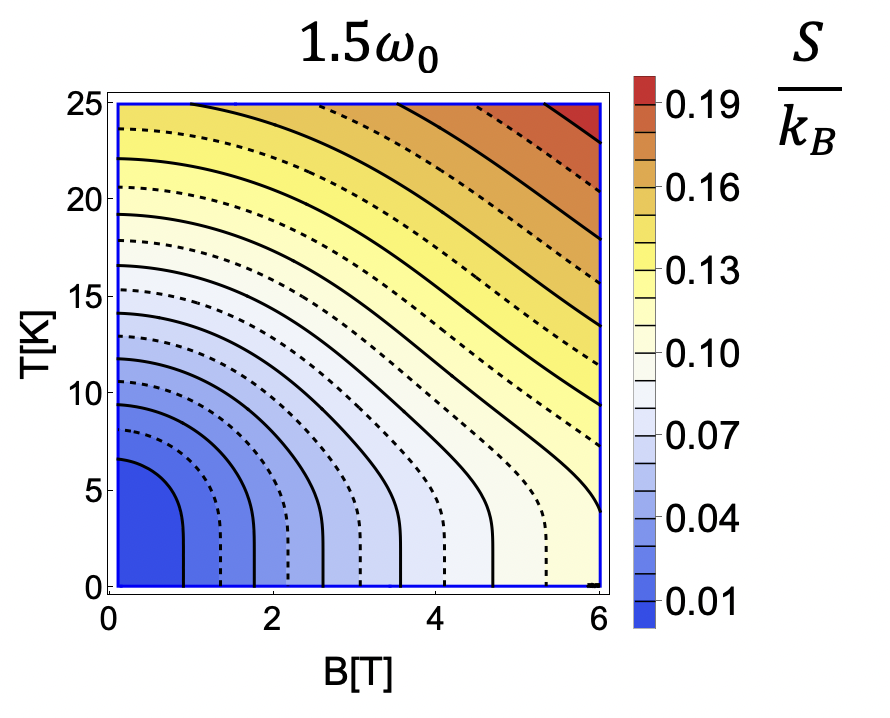}
  \caption{}
  \label{fig:sfig3}
\end{subfigure}
\begin{subfigure}{.5\textwidth}
  \centering
  \includegraphics[width=0.8\linewidth]{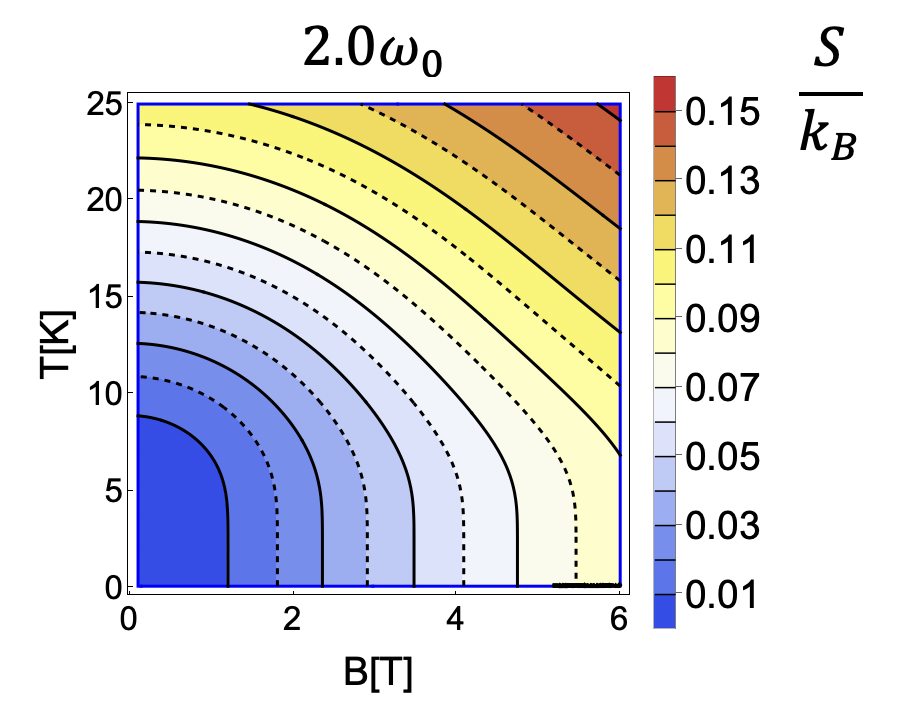}
  \caption{}
  \label{fig:sfig4}
\end{subfigure}
\caption{Contour map showing constant entropy curves as a function of temperature, $T$, (in Kelvin) and the external magnetic field $B$ (in Tesla) for different values of the geometric confinement with $\omega_{0}=2.67$ THz.}
\label{scnt}
\end{figure}

\begin{figure}  
\center
\includegraphics[width=0.7 \textwidth]{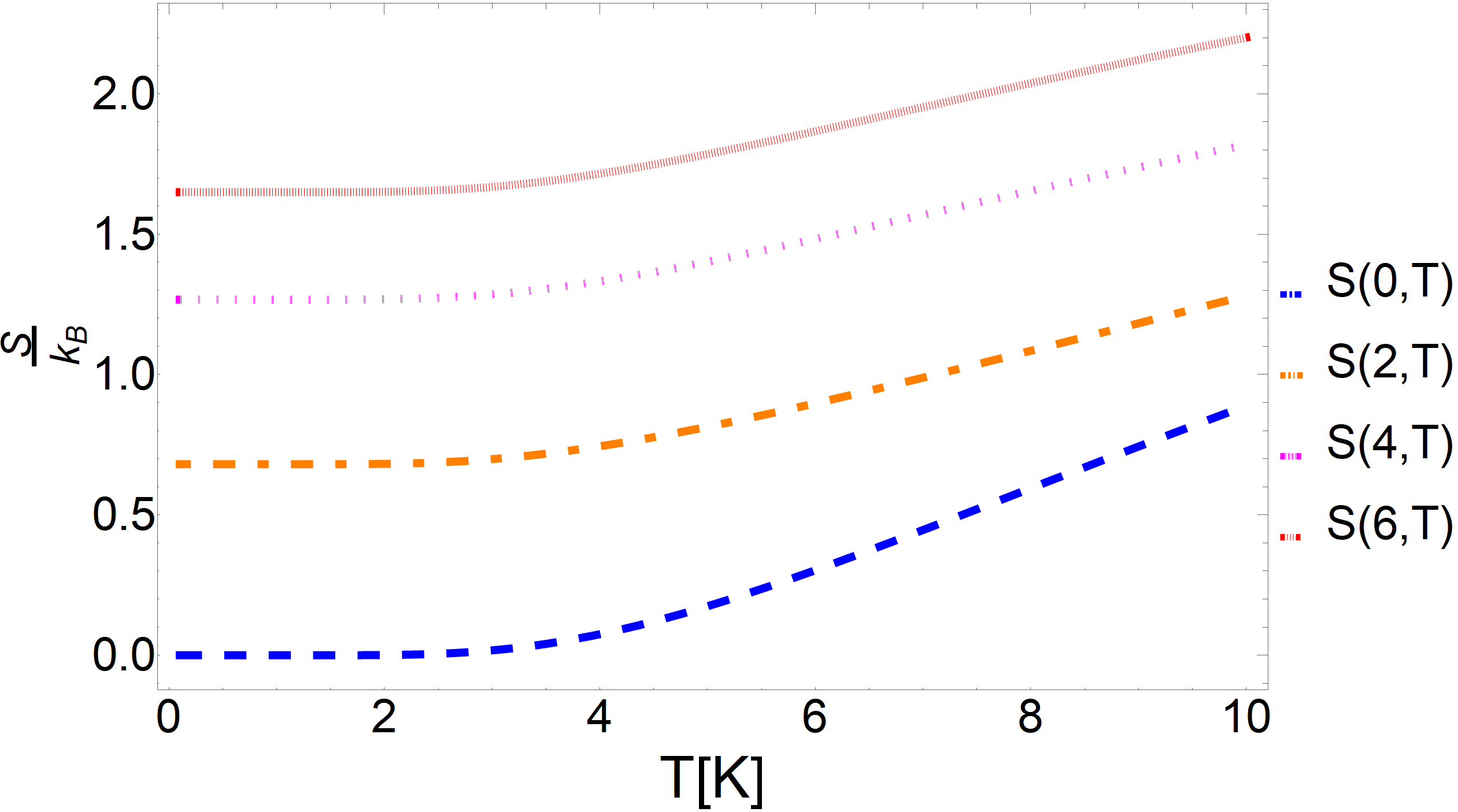}
 \caption{Entropy as a function of temperature for magnetic field strengths of $B=0$ T (blue), $B=2$ T (orange), $B=4$ T (magenta) and $B=6$ T (red).}
 \label{entropyT}
\end{figure}

While these thermodynamic quantities are widely discussed in the literature, here we pay special attention to the conditions under which the entropy remains constant, as this will prove a key aspect of our analysis of the cycle performance. With this in mind, we seek a relationship between the temperature and the external magnetic field that holds the entropy constant. In Fig. \ref{scnt} we plot the isentropic trajectories as a function of the temperature and magnetic field. From this plot, we see that temperature and magnetic field are inversely related, i.e., when we increase the magnetic field, the temperature must decreases if we wish to hold the entropy constant. In Fig. \ref{entropyT}, we plot the entropy as a function of temperature for a range of magnetic field strengths. From this plot, we observe that, at low temperatures, there is a large region where the entropy remains almost constant. In this region, the entropy depends only on the partition function degeneracy term, which is proportional to the value of the external field. Thus, to observe changes in entropy, a considerable increase in the temperature is required. This results in the near-vertical isentropic lines observed at low temperatures in Fig. \ref{scnt}. 

\section{The endoreversible Otto cycle}
\label{Sec3}

\begin{figure}[!ht]  
\center
\includegraphics[width=0.7 \textwidth]{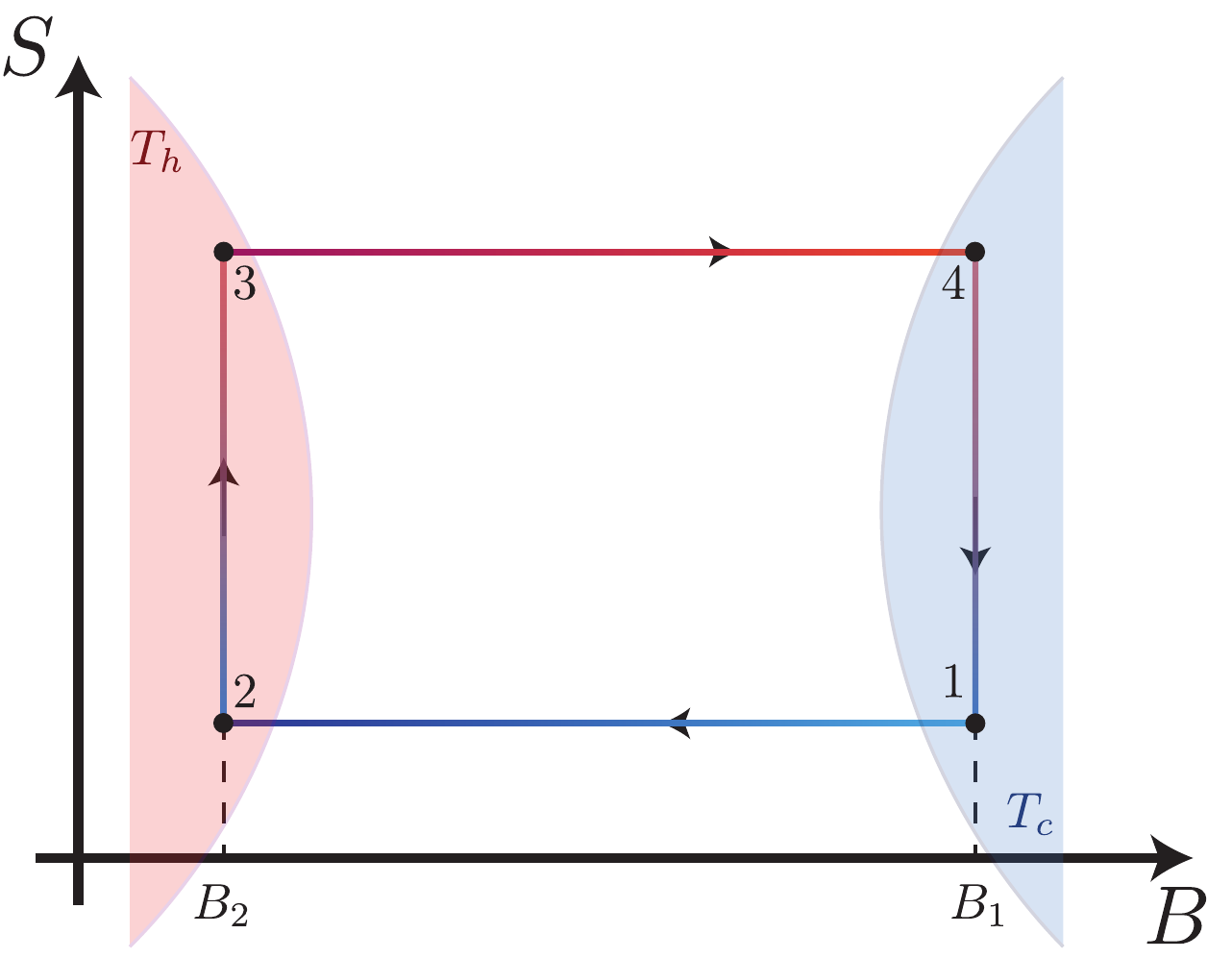}
 \caption{Entropy versus external field diagram for the endoreversible Otto Cycle. Note that the system is only in contact with the thermal reservoirs during the isochoric (vertical) strokes. Under the assumptions of endoreversibility, the working substance does not fully thermalize to the temperatures of the hot and cold reservoirs, $T_{h}$ and $T_{c}$, at points $\mathrm{3}$ and $\mathrm{1}$. 
 }
 \label{OttoCycle}
\end{figure}

The Otto cycle sketched on an entropy-magnetic field diagram in Fig.~\ref{OttoCycle}, consists of four strokes: two isentropic (horizontal lines) processes and two isochoric processes (vertical lines). During the first process ($1\rightarrow 2$) the working substance is disconnected from the thermal reservoir, and the external magnetic field is changed adiabatically from $B_1$ to $B_2$. Since $B_1<B_2$ the internal energy of the working medium increases during this process, despite the fact that the working medium temperature decreases. Thus, this stroke is properly classified as an isentropic compression in which the magnetic field plays a role analogous to the inverse of volume. The next process ($2\rightarrow 3$) is an isochoric heating stroke during which the working substance is put in contact with a thermal reservoir at temperature $T_h$ and allowed to exchange heat while the magnetic field remains constant. The next stroke ($3\rightarrow 4$) is accomplished by disconnecting the working substance from the thermal reservoir and changing the magnetic field adiabatically from $B_2$ back to $B_1$. As the internal energy of the working medium decreases during this stroke, despite increasing in temperature, it corresponds to an isentropic expansion. Finally, the last process ($4\rightarrow 1$) is an isochoric cooling stroke, during which the working substance is put in thermal contact with a reservoir at temperature $T_c < T_h$ and allowed to exchange heat while the magnetic field is again held constant. We note that the inverted behavior of the temperature during the compression and expansion strokes (in comparison to the typical Otto cycle) arises due to inverse relationship between the temperature and magnetic field necessary to maintain constant entropy, as observed in the negative slope of the isentropic curves in Fig. \ref{scnt}. 

The fact that the temperature decreases during the isentropic compression stroke can have negative impacts on the cycle performance, especially in the parameter regions where the decrease is significant, such as the low-temperature, low-magnetic field regimes seen in Fig. \ref{scnt}. In these regimes the isentropic curves become nearly vertical in temperature, indicating that a small increase in the magnetic field must correspond to a large decrease in the temperature if the compression stroke is to remain isentropic. The low temperature of the working medium will then result in a large amount of heat being absorbed from the hot reservoir during the subsequent heating stroke. Since the efficiency is measured by the ratio of total work to heat absorbed from the hot reservoir, a cycle implemented in this region will have a significantly reduced efficiency.

Thermodynamically, the cycle is characterized by the temperatures of the two thermal reservoirs ($T_h$ and $T_c$) and by the initial and final values of the magnetic field ($B_1$ and $B_2$). Under the assumption of endoreversiblity we note that the working medium is assumed to always be in a state of local equilibrium with a well-defined temperature but that it never fully thermalizes with the reservoirs. Following the analysis established in Ref. \cite{Deffner2018Entropy}, we determine the heat exchanged with the reservoirs during the isochoric heating stroke ($2 \rightarrow 3$) using, 
\begin{equation}
\label{qinendo}
Q_{\mathrm{in}}=U_{\mathrm{3}}(T_{3},B_{2})-U_{\mathrm{2}}(T_{\mathrm{2}},B_{2}),
\end{equation}
where we note $T_3, T_2 \neq T_h$. The temperatures $T_{2}$ and $T_{3}$ satisfy the following conditions,
\begin{equation}
    T(0)= T_2 \quad \mathrm{and} \quad T(\tau_{h})= T_{3} \quad \mathrm{with} \quad T_2 < T_{3} \leq T_{h},
\end{equation}
where $\tau_{h}$ is the duration of the heating stroke. We can explicitly model the temperature change from $T_2$ to $T_{3}$ by applying Fourier's law of heat conduction,
\begin{equation}
\label{isochoricheatingdiff}
    \frac{d T}{d t} =-\alpha_{h}\left(T(t)- T_{h}\right),
\end{equation}
where $\alpha_{h}$ is a constant that depends on the working medium's thermal conductivity and heat capacity. Solving Eq.~(\ref{isochoricheatingdiff}) results in, 
\begin{equation}
\label{eq:isoheating}
    T_{3} - T_{h}= (T_2 - T_{h}) e^{-\alpha_{h}\tau_{h}}.
\end{equation}

The isentropic expansion stroke (from $3 \rightarrow 4$) is performed identically to the case of a quasistatic cycle. Since the working medium is decoupled from the thermal reservoirs during this stroke, the work is determined entirely from the change in internal energy,
\begin{equation}
	\label{eq:Wexp}
	W_{\mathrm{exp}} = U_4(T_4, B_{1}) - U_{3}(T_3, B_{2}).
\end{equation} 

The isochoric cooling stroke ($4 \rightarrow 1$) can be modeled identically to the heating stroke. The heat exchanged with the cold reservoir is
\begin{equation}
\label{qoutendo}
Q_{\mathrm{out}}=U_{1}(T_{1},B_{1})-U_{4}(T_4,B_1),
\end{equation}
where $T_{1}$ and $T_4$ satisfy the conditions
\begin{equation}
    T(0)= T_4 \quad \mathrm{and} \quad T(\tau_{c})= T_{1} \quad \mathrm{with} \quad T_4 > T_{1} \geq T_{c}.
\end{equation}
As in the heating stroke, the temperature change can be modeled by Fourier's law, 
\begin{equation}
\label{isochoriccoolingdiff}
    \frac{d T}{d t} =-\alpha_{c}\left(T(t)- T_{c}\right),
\end{equation}
By solving Eq.~(\ref{isochoriccoolingdiff}) we obtain
\begin{equation}
\label{eq:isocooling}
    T_{1} - T_{c} =\left(T_4 - T_{c}\right)e^{-\alpha_{l}\tau_{c}}.
\end{equation}

Finally, the work done during the adiabatic compression stroke ($1\rightarrow 2$) can be found from the change in internal energy,
\begin{equation}
	\label{eq:Wcomp}
	W_{\mathrm{comp}} = U_{2}(T_2, B_{2}) - U_{1}(T_1, B_{1}).
\end{equation} 

The efficiency of the engine can then be found from the ratio of the total work and the heat exchanged with the hot reservoir,
\begin{equation}
	\label{eq:eff}
	\eta = -\frac{W_{\mathrm{comp}}+W_{\mathrm{exp}}}{Q_{\mathrm{in}}}.
\end{equation}
The power output is given by the ratio of the total work to the cycle duration,
\begin{equation}
	\label{eq:P}
	P = -\frac{W_{\mathrm{comp}}+W_{\mathrm{exp}}}{\gamma (\tau_h + \tau_c)},
\end{equation}
with $\gamma>1$ is a multiplicative factor that incorporates the duration of the isentropic strokes \cite{Deffner2018Entropy}.

As the entropy remains constant during the isentropic process, we can obtain a relationship between $T$ and $B$ from the condition $dS(T,B)=0$. This first-order differential equation is given by,
\begin{equation}
\label{differential}
    \frac{d B}{dT}=-\frac{\left(\frac{\partial S}{\partial T}\right)_{B}}{\left(\frac{\partial S}{\partial B}\right)_{T}}.
\end{equation}

For the case of a Fock-Darwin model quantum dot, even though the analytic form of the partition function is known, as given in Eq.~(\ref{partitionfunctionspin}), this equation does not yield a straightforward analytical solution. Nevertheless, it can be solved  numerically, as was done to determine the constant entropy curves in Fig. \ref{scnt}. In addition, it is useful to define the "magnetic length" as,
\begin{equation}
l_{B}=\sqrt{\frac{\hbar}{eB}}.
\end{equation}
By taking the ratio of the magnetic lengths, 
\begin{equation}
r=\frac{l_{B_{2}}}{l_{B_{1}}}=\sqrt{\frac{B_{1}}{B_{2}}},
\label{eq21}
\end{equation}
we obtain a quantity analogous to the compression ratio for the classical Otto cycle.

In the limit of $\tau_{h},\tau_{c}\rightarrow\infty$, the working medium will fully thermalize and we expect to recover the quasistatic cycle behavior. We see that in this limit,
\begin{equation}
\begin{split}
\lim_{\tau_{h}\rightarrow \infty}    \left( T_{3} - T_{h}= (T_2 - T_{h}) e^{-\alpha_{h}\tau_{h}}\right) \rightarrow T_{3} = T_{h},  \\
\lim_{\tau_{c}\rightarrow \infty}     \left( T_{1} - T_{c} =\left(T_4 - T_{c}\right)e^{-\alpha_{c}\tau_{c}} \right)  \rightarrow T_{1} = T_{c}.
\end{split}
\end{equation}
However, this limit also results in vanishing power output, as seen from Eq. ~(\ref{eq:P}). 

 \section{Results and Discussions}
 \label{Sec4}

\begin{figure}
\begin{subfigure}{.5\textwidth}
  \centering
  \includegraphics[width=.9\linewidth]{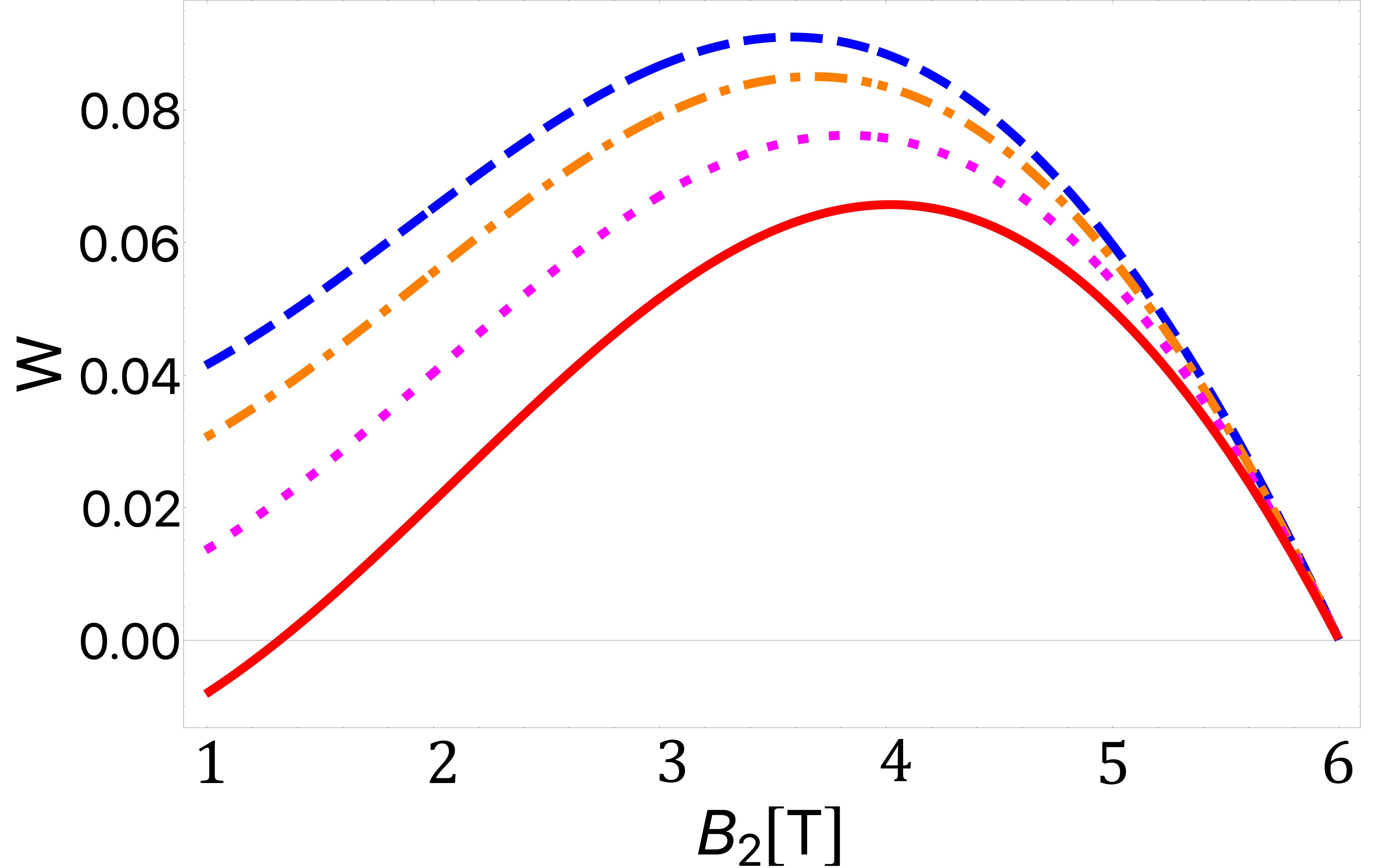}
  \caption{}
  \label{fig:work1}
\end{subfigure}
\begin{subfigure}{.5\textwidth}
  \centering
  \includegraphics[width=.9\linewidth]{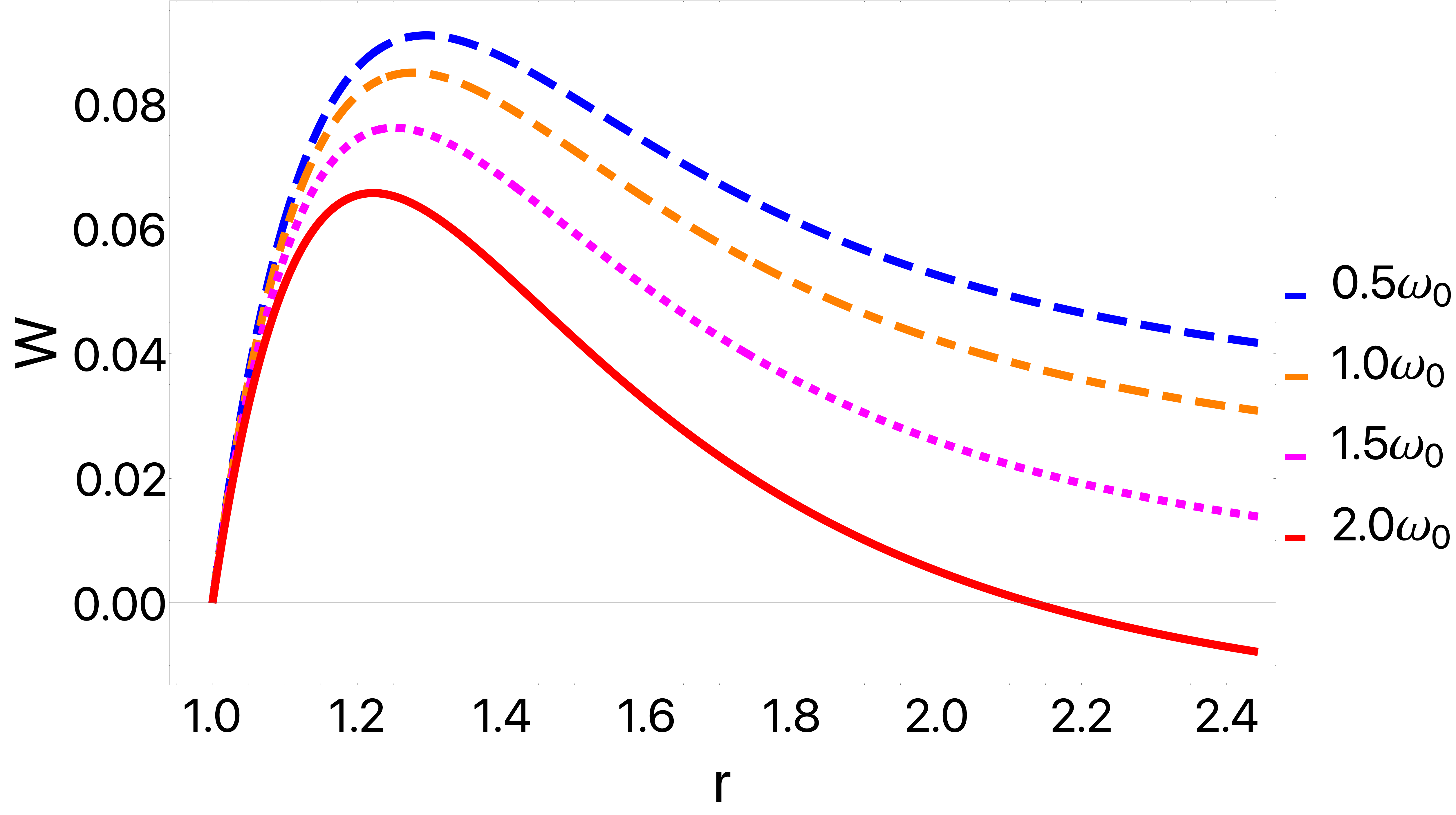}
  \caption{}
  \label{fig:work2}
\end{subfigure}
\caption{Total work as a function of (a) the external field and (b) the compression ratio for geometric confinement frequencies of $0.5 \omega_0$ (blue), $\omega_0$ (orange), $1.5\omega_0$ (magenta) and $2\omega_0$ (red). We have taken $T_{c}=13$ K, $T_{h}=25$ K, and $B_{1}= 1$ T. Note the clear decrease in the net work with increasing values of the dot confinement. In particular, for these values a transition to negative net work is observed for the case of $2\omega_{0}$ (red curve). At this transition point the cycle switches from behaving as an engine to behaving as a refrigerator.}
\label{fig:work1def}
\end{figure}

\begin{figure}[ht]
	\centering
	\includegraphics[width=0.6\textwidth]{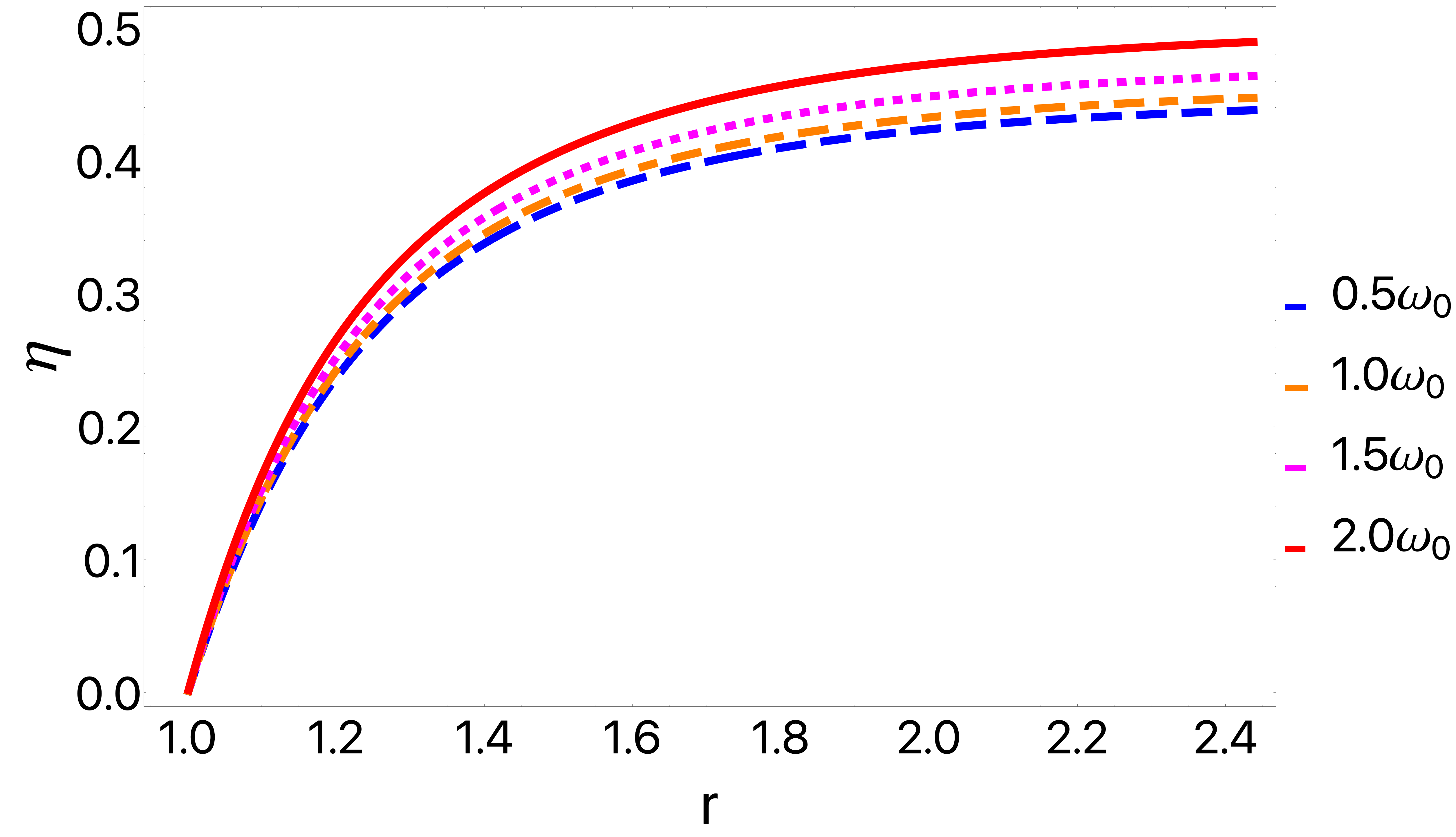}
	\caption{Efficiency as a function of $r$ for geometric confinement frequencies of $0.5 \omega_0$ (blue), $\omega_0$ (orange), $1.5\omega_0$ (magenta) and $2\omega_0$ (red). We have taken $T_{c}=13$ K and $T_{h}=25$ K. We observe that the system's efficiency increases as we increase the geometric confinement.}
		\label{eff1}
\end{figure}

We begin by considering a specific example with bath temperatures of $T_{c}=13$ K and $T_{h}=25$ K in order to illustrate the cycle performance. These results are presented in Fig.~\ref{fig:work1def}, for the work extracted and in Fig.~\ref{eff1} for the efficiency. In Fig.~\ref{fig:work1def}, we observe that as the parabolic trap parameter $\omega_{0}$ decreases, the amount of work extracted increases. Conversely, the efficiency presented in Fig.~\ref{eff1} follows the opposite behavior, growing as $\omega_{0}$ increases. This comparison demonstrates a clear trade-off between work and efficiency. In addition, we observe in Fig.~\ref{fig:work1def} that there exists a region in which the total work becomes negative for the case of $2\omega_{0}$, indicating that our cycle is no longer behaving as an engine in this regime. With this in mind, we next turn to explore the parameter regimes where the cycle functions as different types of thermal machines.

In general, there exist four possible types of thermal machines, corresponding to all possible combinations of heat and work flow consistent with the first and second laws of thermodynamics. An engine corresponds to positive work output, along with heat flow from the hot bath into the working medium and from the working medium into the cold bath. A refrigerator corresponds to negative work output, along with heat flow from the cold bath into the working medium and from the working medium into the hot bath. A heater corresponds to negative work output and heat flow from the working medium into both baths. Finally, an accelerator corresponds to negative work output along with heat flow from the hot bath into the working medium and from the working medium into the cold bath.

\begin{figure}
\begin{subfigure}{.5\textwidth}
  \centering
  \includegraphics[width=.6\linewidth]{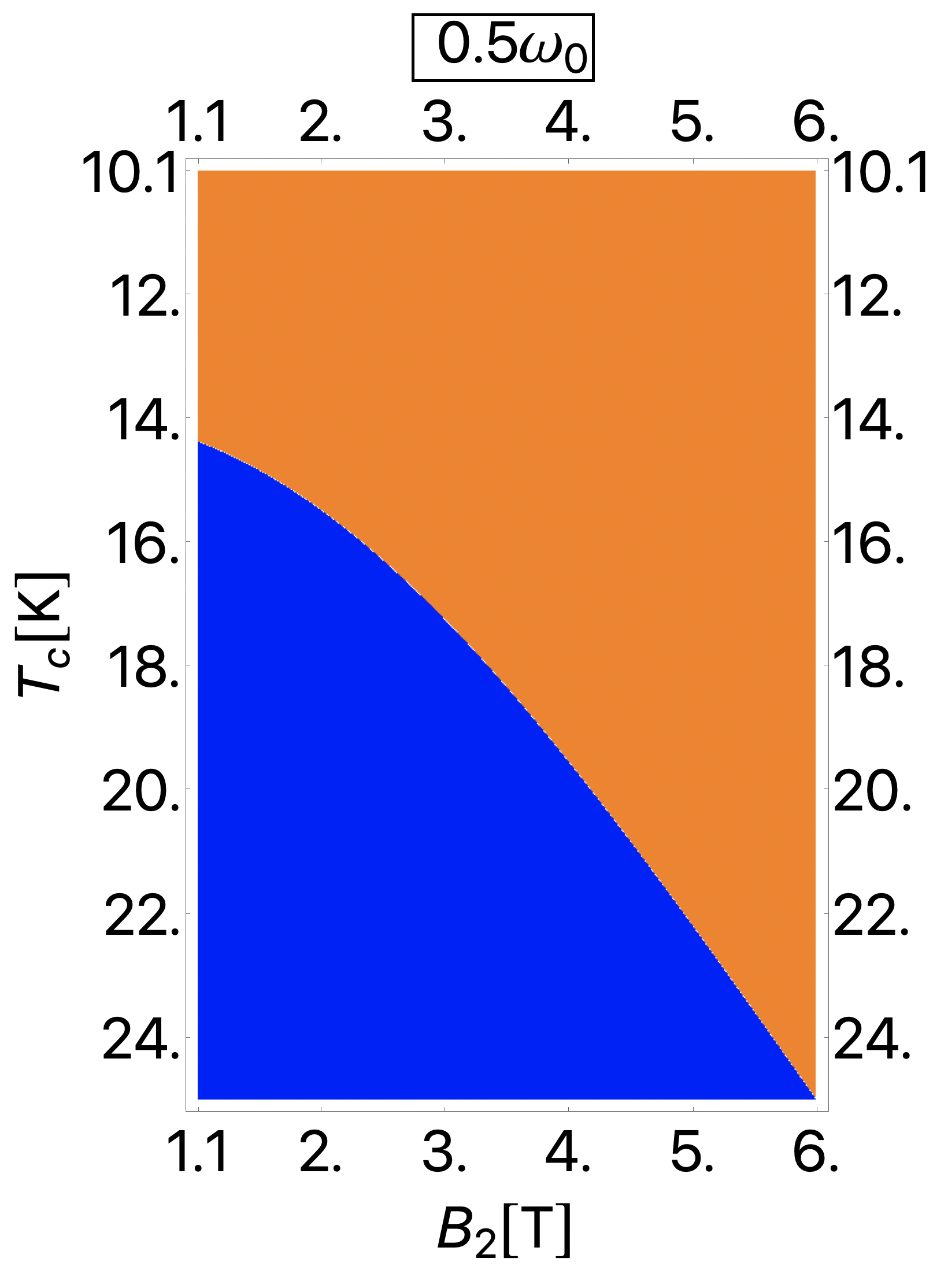}
  \caption{}
  \label{fig:engine0.5}
\end{subfigure}
\begin{subfigure}{.5\textwidth}
  \centering
  \includegraphics[width=.8\linewidth]{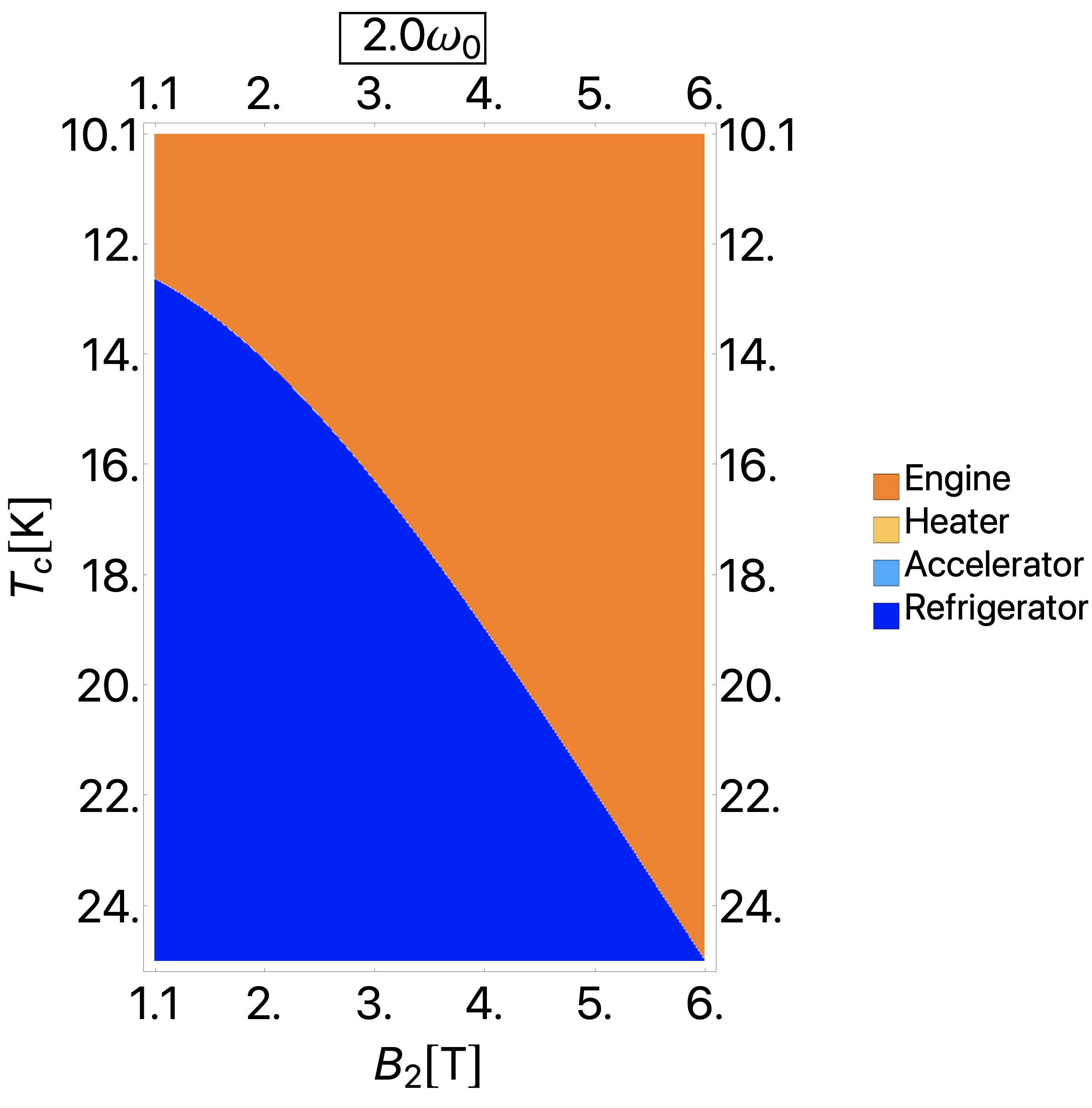}
  \caption{}
  \label{fig:engine2.0}
\end{subfigure}
\caption{Cycle behavior as a function of the external magnetic field strength and cold bath temperature for geometric confinement of (a) 0.5$\omega_{0}$ and (b) $2\omega_{0}$. Note the increased size of the refrigerator region as the value of the parabolic trap frequency increases.}
\label{fig:enginerefrigerator}
\end{figure}

By numerical calculation, we determine the results of Eqs. (\ref{qinendo}), (\ref{eq:Wexp}), (\ref{qoutendo}), and (\ref{eq:Wcomp}) across the parameter space. By comparing the signs of the resulting heat and work flows, we determine the regions where the cycle behaves as an engine, refrigerator, heater, or accelerator.

Figure~\ref{fig:enginerefrigerator} shows the regions where the cycle operates as an engine or as a refrigerator as a function of the cold bath temperature and the external magnetic field for two different values of $\omega_{0}$ while holding the temperature of the hot reservoir constant at $T_h=25$K. We observe that by increasing the frequency of geometric confinement, the region in which the cycle operates as an engine is reduced. This behavior can be understood from Fig.~\ref{fig:work1def}, where it is observed that as the trap frequency increases, the extracted work decreases, resulting in a transition into the regime of negative work. From Fig.~\ref{fig:enginerefrigerator}, we see that increasing the value of $T_{c}$ increases the value of the magnetic field at which the transition to the refrigerator regime occurs. Recalling Fig.~\ref{fig:work1def}b, increasing $T_{c}$ corresponds to shifting the peak to the left, and thus also shifting to the left the critical point at which the transition to the negative work regime occurs while maintaining the same qualitative behavior.

  \begin{figure}
	\centering
	\includegraphics[width=0.8\textwidth]{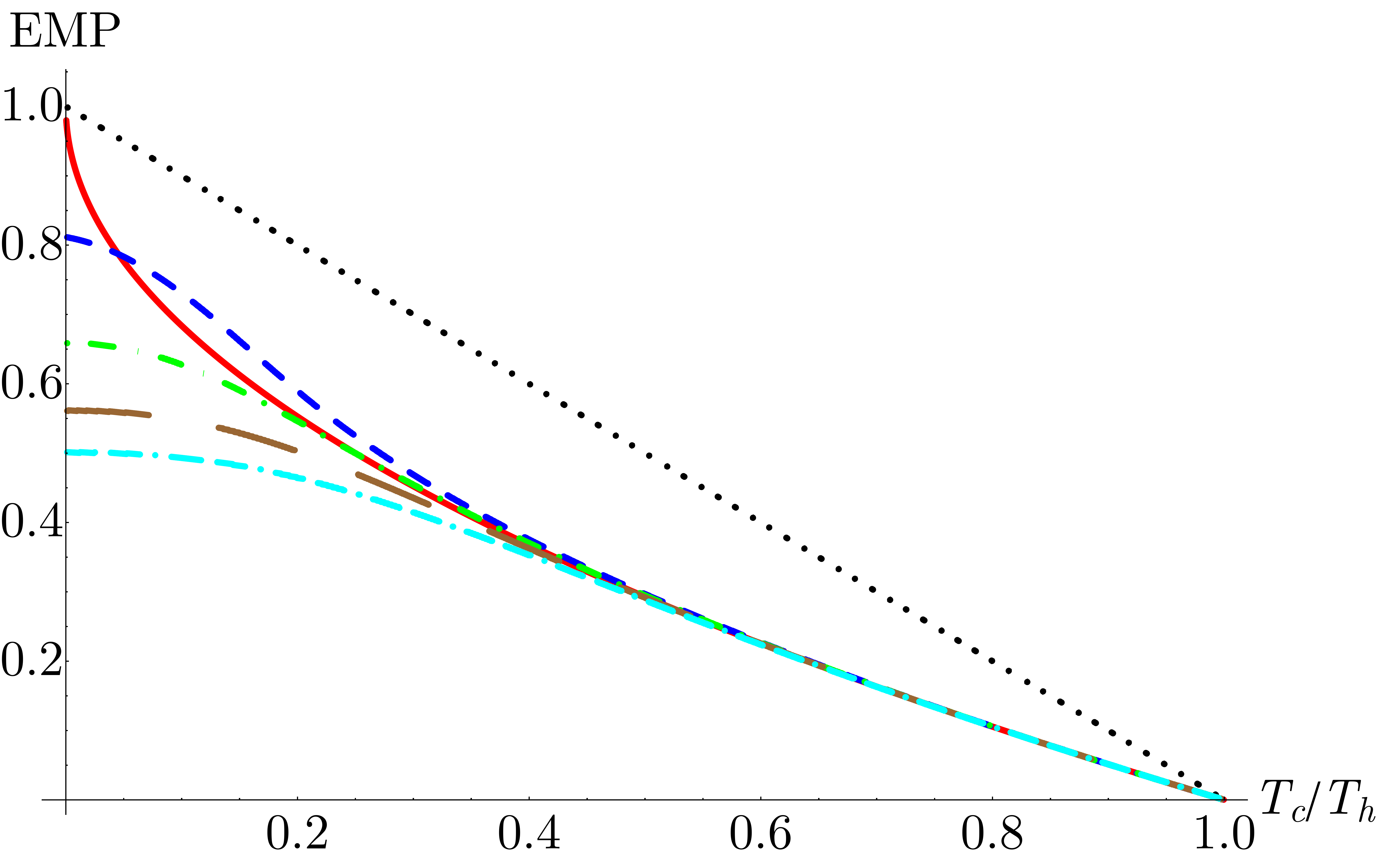}
	\caption{Efficiency at maximum power as a function of the bath temperature ratio for $0.5 \omega_0$ (blue, short dashed), $ 1.0 \omega_0 $ (green, dot-dashed), $ 1.5 \omega_0 $ (brown, long dashed), and $ 2.0 \omega_0 =$ (cyan, dot-dash-dashed). The  Carnot (black, dotted) and Curzon-Ahlborn (red, solid) efficiencies are provided for comparison. We observe that the efficiency at maximum power exceeds CA for lower values of dot confinement frequency at low bath temperature ratios. Parameters are $T_{h}$=25 K and magnetic field $B_{h}=$ 12 T. }
    \label{EMP}
\end{figure}

To numerically determine the EMP, we fix the temperature of the hot bath at $T_h=25$K and the high magnetic field value at $B_{h}=$ 12 T. We then vary the cold bath temperature across the range $T_c\in[0\textrm{K},25\textrm{K}]$ and the low magnetic field value from $B_l\in[0\textrm{T},12\textrm{T}]$. This range of parameter space is selected to ensure that the peak in the work as a function of the compression ratio (see Fig. \ref{fig:work2}) is captured for all combinations of bath temperatures. For each bath temperature ratio we scan the parameter space to determine the magnetic field strengths that produce the maximum power output and then determine the efficiency at that maximum power. In Fig.~\ref{EMP} we plot the EMP as a function of the bath temperature ratio $T_{c}/T_{h}$ for different values of $\omega_{0}$. These EMPs are then compared with the CA efficiency, found in Eq.~(\ref{CAeff}).

From Fig.~\ref{EMP}, we can see that the geometric confinement frequency has a significant impact on the EMP at low bath temperature ratios. Lower values of the confinement frequency yield higher EMPs. In particular, for the case of $\omega_{0} = 0.5$ we see that there exists a range of bath temperatures where the EMP exceeds the CA efficiency. At higher bath temepratures, we see the EMP converges towards the CA efficiency for all values of the confinement frequency, as is expected in the high-temeprature, classical limit.

Previous results have shown that the EMP of an endoreversible quantum Otto engine with a pure harmonic oscillator as the working medium exceeds the CA efficiency \cite{Deffner2018Entropy}. As the partition function in Eq.~(\ref{partitionfunctionspin}) simplifies to that of a two-dimensional isotropic harmonic oscillator in the limit $\omega_{0} \rightarrow 0$ \cite{KumarPhysRevE2009}. finding that low $\omega_{0}$ leads to a region where the CA efficiency can be exceed is consistent. Here we have further demonstrated that the addition of geometric confinement has a strictly detrimental impact on the engine performance.   

An additional feature of note is the asymptotic flattening of the EMP as the bath temperature ratio approaches zero. This behavior again arises from the structure of the isentropic curves seen in Fig.~\ref{scnt}. As the entropy is nearly independent of temperature in the low temperature regime, the value of the compression ratio that maximizes the power output (and thus the value of the EMP) will not change significantly until the system moves out of this regime. This explains why the asymptotic behavior is especially pronounced at high values of the confinement frequency, which display larger ranges of temperatures under which the isentropic curves are nearly vertical with respect to the magnetic field.


\section{Conclusions}
\label{Sec5}

In this work, we have analyzed the performance of an endoreversible Otto cycle using a quantum dot-trapped electron in the presence of an external magnetic field as a working substance. Modeling our working substance using the Fock-Darwin Hamiltonian, we have found that the net work, power, and efficiency depend strongly on the intensity of the parabolic trapping frequency, as well as influencing whether the cycle behaves as either an engine or a refrigerator. In particular, we found that larger values of the trapping frequency lead to increased efficiency but reduced work, power, and EMP. Despite this, we demonstrated that there exists a parameter regime in which the cycle EMP exceeds the Curzon-Ahlborn efficiency for small values of the trapping frequency. Furthermore, we found that the  strongly nonlinear temperature dependence of the entropy for the Fock-Darwin model quantum dot has distinct signatures in the engine performance, namely an asymptotic behavior of the EMP at low temepratures.


\acknowledgments{F.J.P. acknowledges financial support from ANID Fondecyt, Iniciaci\'on en Investigaci\'on 2020 grant No. 11200032, ANID Fondecyt grant No. 1210312, “Millennium Nucleus in NanoBioPhysics” project NNBP NCN2021 \textunderscore 021. N.M.M. acknowledges support from AFOSR (FA2386-21-1-4081, FA9550-19-1-0272, FA9550-23-1-0034) and ARO (W911NF2210247, W911NF2010013). F.A.-A. acknowledges financial support from ANID Subvenci\'on a la Instalaci\'on en la Academia SA77210018 and ANID Proyecto Basal AFB 220001. P.V. acknowledges support from ANID Fondecyt grant No. 1210312 and ANID PIA/Basal grant No. AFB 220001.}

\authorcontributions{F.J.P. and N.M.M. have carried out the calculations and analysis detailed in the manuscript. F.J.P. wrote the first version of the manuscript. N.M.M., F.A.A., D.O., and P.V. have contributed to discussions and writing the final version of the manuscript.  All authors have read and approved the final manuscript.}

\conflictsofinterest{The authors declare no conflict of interest.} 


\appendixtitles{no} 
\appendixsections{multiple} 


\reftitle{References}

\end{document}